\documentclass[conference]{IEEEtran}
\usepackage[compatibility=false]{caption}
\usepackage[colorlinks=true, linkcolor=blue, citecolor=red, urlcolor=blue]{hyperref}
\IEEEoverridecommandlockouts
\usepackage{cite}
\usepackage{caption}
\usepackage{amsmath,amssymb,amsfonts}
\usepackage{algorithmic}
\usepackage{graphicx}
\usepackage{textcomp}
\usepackage{xcolor}
\def\BibTeX{{\rm B\kern-.05em{\sc i\kern-.025em b}\kern-.08em
    T\kern-.1667em\lower.7ex\hbox{E}\kern-.125emX}}

\begin{document}

\title{Progress in photonic technologies for next-generation astrophotonic instruments}

%Author Details

\author{\IEEEauthorblockN{ Ahmed Shadman Alam}
\IEEEauthorblockA{\textit{Department of Electrical and Computer Engineering} \\
\textit{University of Maryland College Park}\\
College Park, MD 20742, USA \\
alam22@umd.edu}

}

\maketitle

\begin{abstract}
The study of integrating photonic devices into astronomical instruments is the primary focus of astrophotonics. The growth in this area of study is relatively recent. Research related to astronomical spectroscopic phenomena has received a lot of attention in recent times. There are several important advantages to integrating photonics technology into an astronomical instrument, such as cost savings because of increased reproducibility, thermal and mechanical stability, and miniaturization. This paper provides a brief review of recent advances in astrophotonics.
\end{abstract}


\section{Introduction}
\begin{figure}[htbp]
\centering % Centers the image
\includegraphics[width=0.5\textwidth]{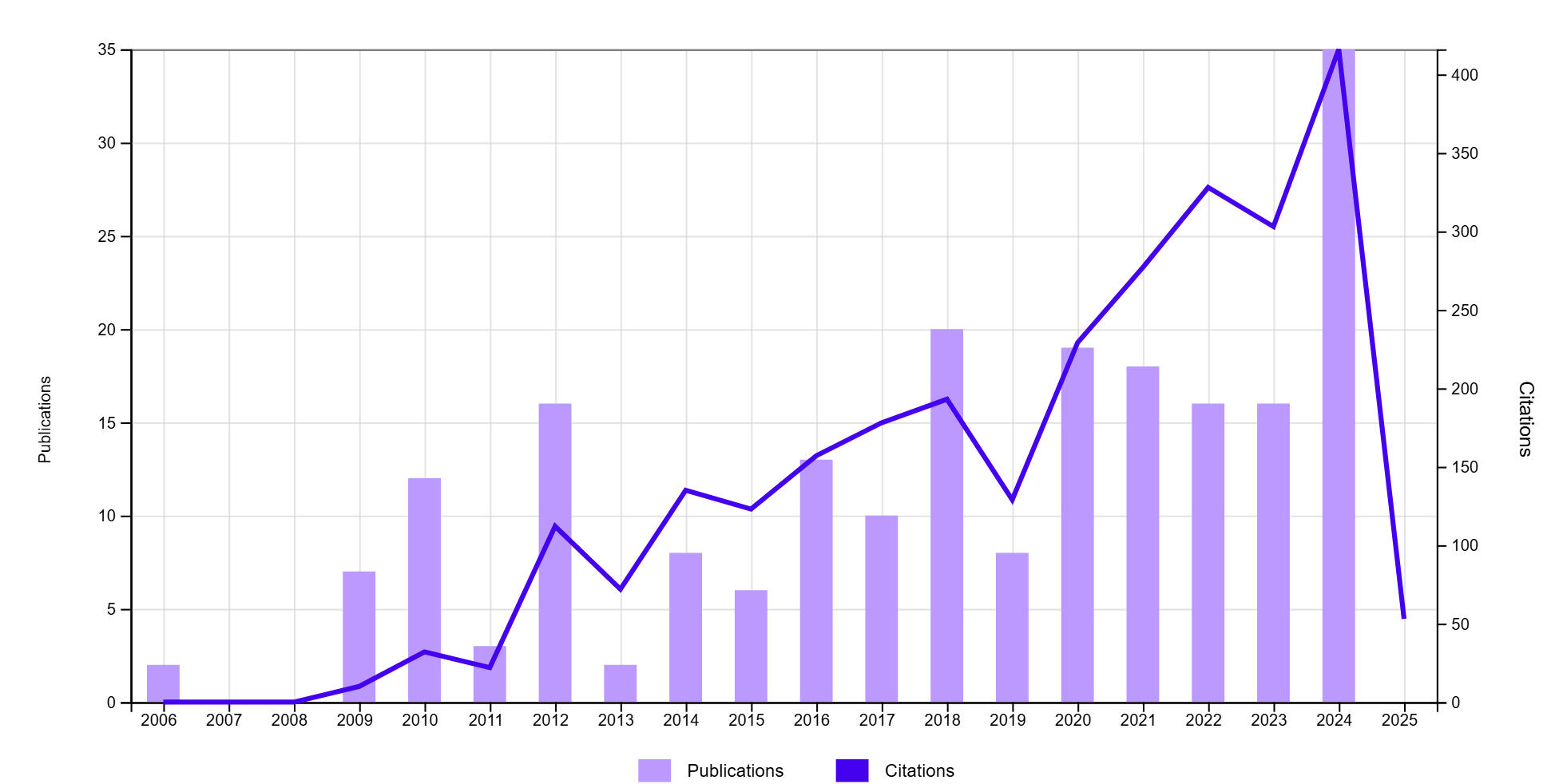} % Ensure the image file path is correct
\caption{Total number of citations to papers in the field of astrophotonics. These data were obtained from the Web of Science interface (\href{https://www.webofscience.com/}{Web of Science}).}
\label{fig1}
\end{figure}

Astronomy is the study of electromagnetic waves from space. Thus, advances in optical technology have always impacted the development of astronomical instruments. A branch of study in optical technology is photonics, which deals with the generation, guiding, and measurement of light waves. The integration of photonics into devices has emerged as a key area of research with the evolution and advancements for wireless and cellular communication technology \cite{18, 16, 17}. The growing importance of photonics has led to its incorporation into astronomical instrumentation. A crucial part of astronomy depends on using light waves to analyze and extract as much information as possible. Future pursuits in astronomy will rely heavily on the development of large telescopes, and photonic technology will play a key role in the innovation of these instruments. It will provide the flexibility for scientists to make new instruments compact and economical. The importance of this field of study has grown exponentially over the last couple of decades, as shown in Fig. \ref{fig1}. This figure illustrates this by plotting the number of papers and citations in this field from 2000 to 2024. Most of the recent work in astrophotonics has been focused on spectrographs. Works on very efficient, ultra-broadband, low-resolution spectrographs \cite{1}, photonic lanterns integrated into telescopes \cite{2}, photonic-lantern-based wavefront sensing, and photonic interferometry are just a few examples of recent advancements in astronomical equipment with the help of photonics integration. This paper reviews recent advancements in the field while also discussing reviews from past decades and evaluating where astrophotonics stands today compared to its predicted trajectory.

% first image

%Section 2
\section{The Science of Photonics}

\subsection{Guiding Light}

One of the fundamental characteristics of a photonic system is the coupling of light into the optical fiber system. To collect light efficiently into the optical system different methods have been utilized over the years. Adaptive optics (AO) is a way that enables us to distort or correct the light waves to compensate for existing wavefront distortions. The wavevector of the light wavefronts must be normal to the fiber input face for light to be coupled to it \cite{2}. The fact that there is a limited amount of light photons to work with requires that instruments built for such astronomical purposes have low insertion losses. Furthermore, total internal reflection occurs because the light travels along the optical axis at angles smaller than the critical angle, resulting in a peak injection angle of $\phi$ given by

\[
n_0 \sin\phi \leq \sqrt{n_{\text{core}}^2 - n_{\text{clad}}^2}
\]
where $n_0 \sin\phi$ is the numerical aperture of the fiber \cite{3}.

\subsection{Photonic Lantern}
\begin{figure}[htbp]
\centering % Centers the image
\includegraphics[width=0.4\textwidth]{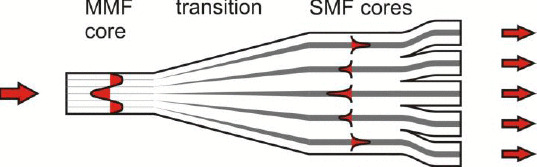} % Ensure the image file path is correct
\caption{Photonic Lantern: A multimode fiber device that is made up of a series of identical single-mode fibers. Adapted from \textit{S. G. Leon-Saval et al.} \cite{5}.}
\label{fig2}
\end{figure}

Although the incorporation of AO is one way of improving the coupling, the practicality of not achieving perfect correction will lead to spatially smeared output beams that will have poor coupling efficiency. Another device that can be used to counteract this is the photonic lantern. Fig. \ref{fig2} shows the diagram of a photonic lantern.

% second image

The lantern has a large input multimode fiber (MMF) device that can be positioned at the focus of a telescope to extract all the light waves. Depending on the number of modes coupled into the lantern, it can be coupled efficiently into different single-mode fiber (SMF) cores. Splitting into SMF waveguides makes the integration of other photonic devices easier. Single-mode operation is very important for accurate measurements of the radial speed of exoplanets. A review carried out by P. Gatkine et al. talks about different methods to couple AO-corrected light on a single-mode fiber \cite{5}.

\subsection{Interferometers}

\begin{figure}[htbp]
\centering % Centers the image
\includegraphics[width=0.25\textwidth]{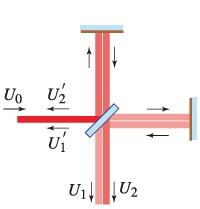}
\caption{Michelson Interferometer: The input wave \( U_0 \) is split into two waves by the beam splitter. After reflecting from the mirrors, they are recombined at the splitter. Adapted from \textit{Fundamentals of Photonics}, Chapter 2 \cite{6}.}
\label{fig3}
\end{figure}

Using instruments known as interferometers is another way to enhance optical waves from space. These optical devices strategically use beam-splitters and mirrors to split and subsequently delay input waves. After that, the delayed waves are recombined using directional couplers or multimode interference couplers, and specific properties like the spatial coherence of the input wave can be described by examining the intensity of the recombined waves. Reproducing high-resolution images and lowering the light from nearby bright stars are specifically necessary for some astronomical research. In some cases these were accomplished by feeding an interferometer with the output from two or more telescopes. Fig. \ref{fig3} illustrates an example of an interferometer showing the placement of its mirrors and beam splitter.

% Section 3

\subsection{Astrophotonic Spectrographs}

Spectroscopy uses the splitting of input waves into different wavelengths by utilizing dispersive optical devices like prisms or gratings \cite{7}. After dispersing the light, it is focused on a surface for detection using a lens. One of the critical parameters of spectroscopy is the resolving power, defined as \( R = \frac{\lambda}{\Delta \lambda} \). This is the measure of how well the device can measure the flux densities at two points that are separated by \( {\Delta \lambda}\).

Since the light source is so weak, low-resolution spectroscopy is a crucial tool for astronomical studies. An example of such low-resolution spectroscopy that enables compact on-chip integration for astronomical applications is the work done by P. Gatkine et al. \cite{1} on the characterization of broadband arrayed waveguide grating (AWG) spectrographs based on low-loss technologies of \(\text{Si}_3\text{N}_4\) and doped \(\text{SiO}_2\). They reported a maximum resolving power of approximately 200, a free spectral range of 200–300\, nm, and a compact form factor of 50–100\,\(\text{mm}^2\). 

Although the doped-\(\text{SiO}_2\) platform-based devices demonstrated a 79\% (1\,dB) peak fiber-to-chip efficiency with negligible polarization dependence, the AWGs based on \(\text{Si}_3\text{N}_4\) exhibited a lower efficiency of around 50\% (3\,dB). Based on their research, this reduced efficiency was primarily attributed to fiber-to-chip coupling loss.

%Section 4

\section{Integration of Photonics in Astronomical Instruments}

\begin{figure}[htbp]
\centering % Centers the image
\includegraphics[width=0.45\textwidth]{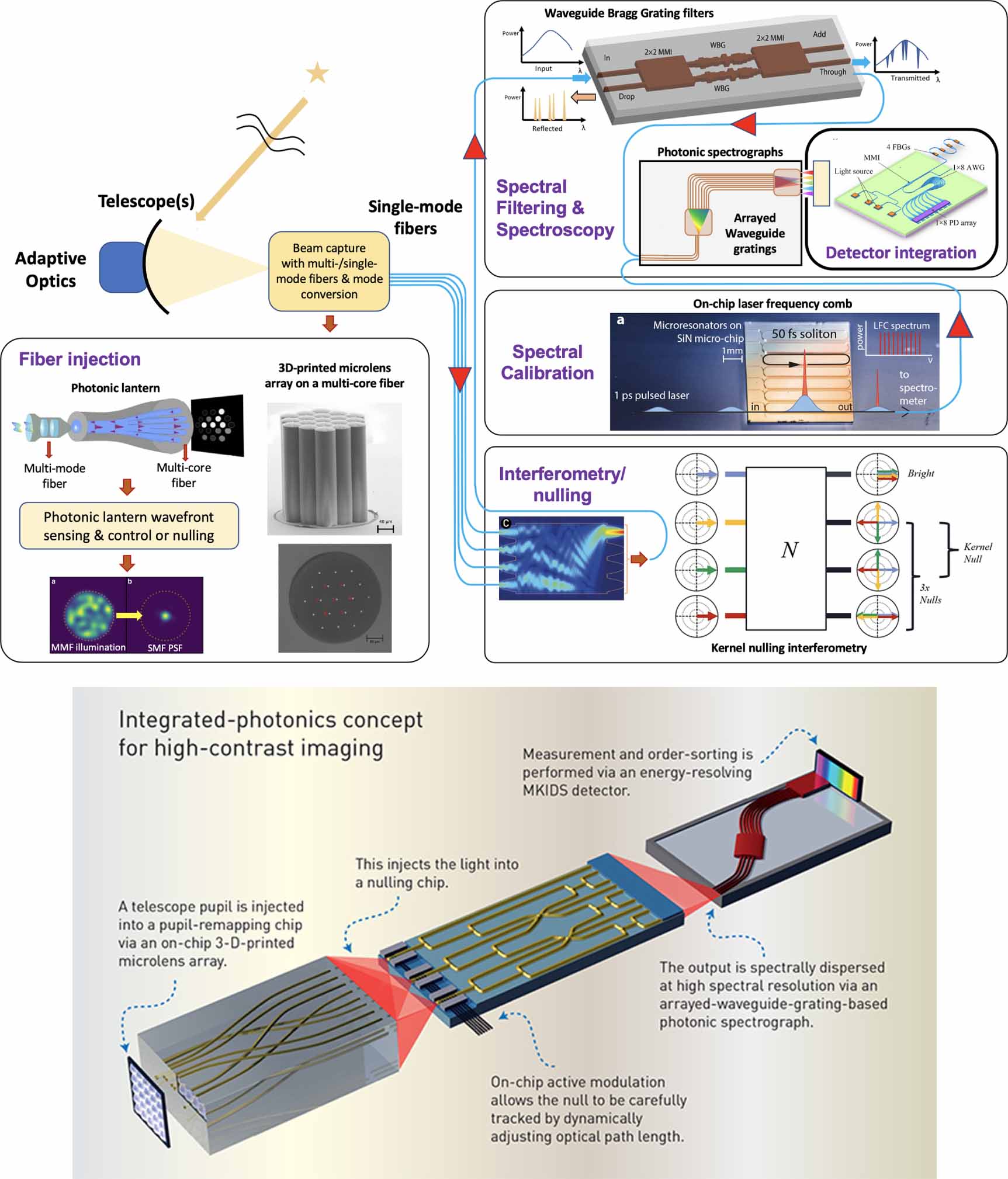} % Ensure the image file path is correct
\caption{Astrophotonics' various subfields and the photonic equipment utilized at different stages of the analysis. Adapted from Jovanovic et al. \cite{9}.}
\label{fig4}
\end{figure}

With the numerous positive aspects of photonics discussed in previous sections, photonics has emerged as an important catalyst in the development of astronomical research. Some of the primary subtopics covered in this paper, along with their present state and potential future developments, are shown in Fig. \ref{fig4}.

\subsection{Interfeometry and Telescopes}
\begin{figure}[htbp]
\centering % Centers the image
\includegraphics[width=0.45\textwidth]{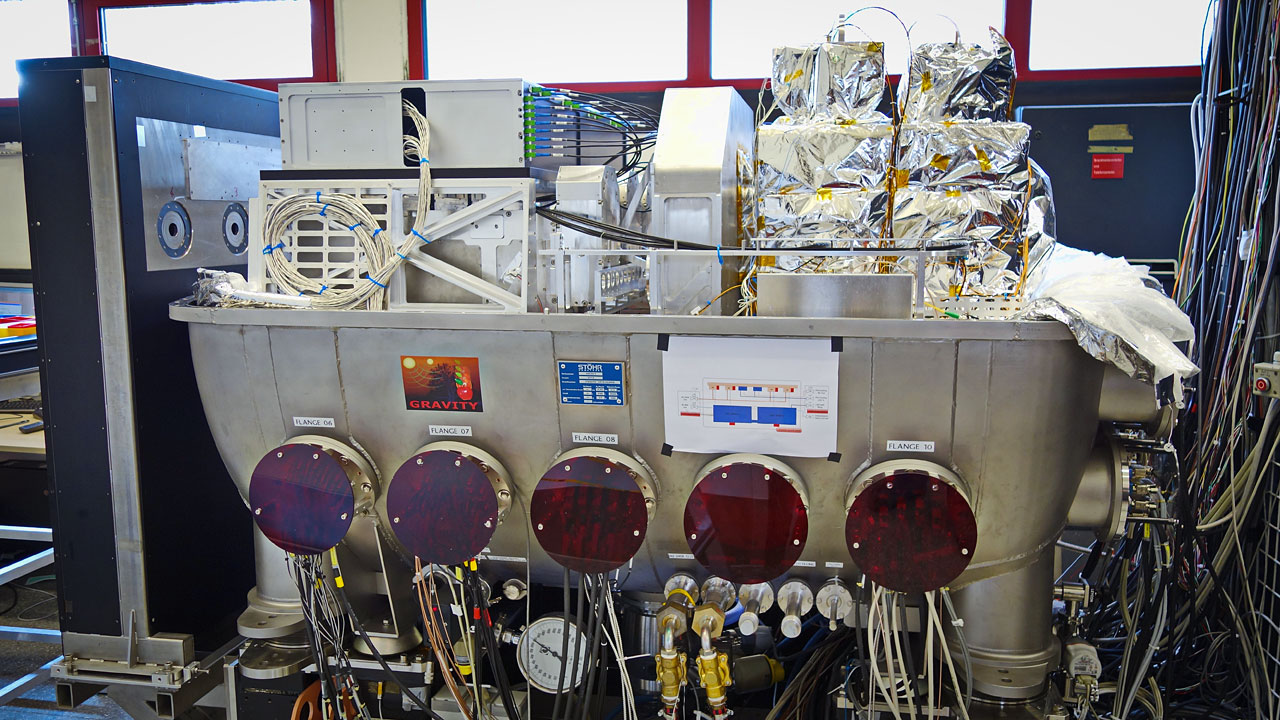}
\caption{GRAVITY is an interferometric instrument that combines beams from eight telescopes to perform phase referencing and interferometric imaging for astronomical purposes. Image taken from \href{https://www.eso.org/sci/facilities/paranal/instruments/gravity/}{www.eso.org}.
}
\label{fig5}
\end{figure}

One significant development in this area was the realization of the GRAVITY instrument (shown in Fig. \ref{fig5}) at the Very Large Telescope Interferometer (VLTI), operated by the European Southern Observatory (ESO). This achievement was made possible by combining several outputs from multiple telescopes, integrated photonics, and free-space optics.

\begin{figure}[htbp]
\centering % Centers the image
\includegraphics[width=0.5\textwidth]{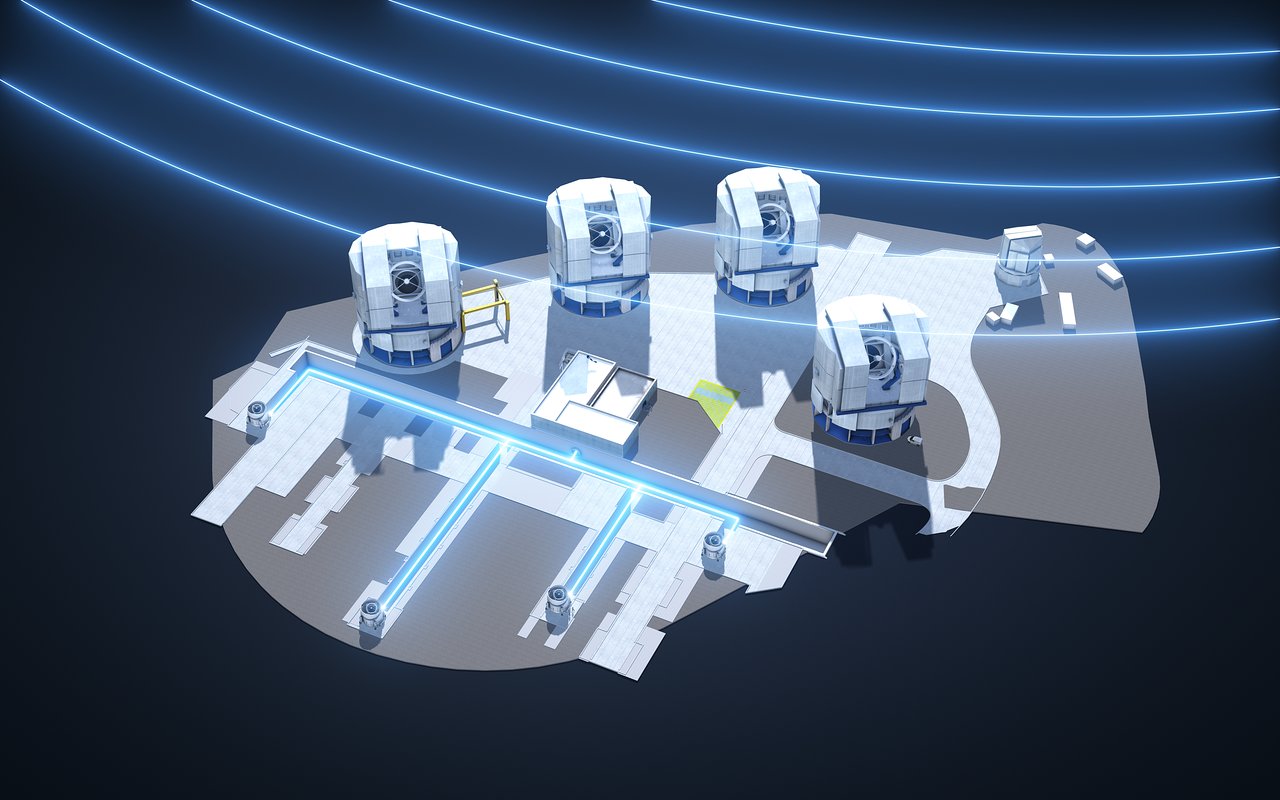}
\caption{The eight VLT telescopes at work. Image taken from \href{https://www.eso.org/sci/facilities/paranal/instruments/gravity/}{www.eso.org}.
}
\label{fig6}
\end{figure}

When combining the telescope beams, it is necessary to compensate for the light beams that arrive at the various telescopes at different times to guarantee accuracy. Four large Unit Telescopes (with a mirror diameter of 8.2 m) and four small Auxiliary Telescopes (with a mirror diameter of 1.8 m) make up the eight telescopes in the VLT. Where rails allow the latter to be moved into various positions. With this configuration, distant objects can be resolved to the same resolution as a telescope with a diameter of 100 meters. Fig. \ref{fig6} shows the telescopes at the VLT.

% The GRAVITY instrument at the VLT is undergoing an upgrade to GRAVITY+. This upgrade will feature improvements such as the use of advanced  AO to mitigate dispersive effects of the Earth's atmosphere and enhance the contrast of observations.

The GRAVITY instrument at the VLT is undergoing an upgrade to GRAVITY+. This upgrade will feature improvements such as the use of advanced  AO to mitigate dispersive effects of the Earth's atmosphere and enhance the contrast of observations.

\subsection{Advancements in Astrophotonic Spectrosopy}

The demand for photonic devices in the telecommunications sector has led to significant technological advancements in recent years. AWGs, or arrayed waveguide gratings, are frequently employed in such applications. Based on a phased-arrayed channel found in radio receivers, the concept of AWGs was initially introduced in 1988 \cite{7}. Photonic AWGs are one of the most explored on-chip dispersion techniques. The feasibility of drawing small nanoscale structures using electron beam lithography has contributed to this advancement. Moreover, \(\text{Si}_3\text{N}_4 / \text{SiO}_2\) has become a go-to platform for low-loss and high-index contrast in photonic devices.

As mentioned in Section II, the research of P. Gatkine et al. has utilized this platform for their broadband low-resolution AWGs \cite{1}. Table \ref{table1} provides a summary of the characteristics of these AWGs that are in different material platforms. 

% Table of summary (Cite 1)
\begin{table}[h]
    \centering
    \renewcommand{\arraystretch}{1} % Adjust row height
    \setlength{\tabcolsep}{2pt} % Adjust column spacing
    \begin{tabular}{|l|c|c|c|}
        \hline
        & \textbf{AWG\#1} & \textbf{AWG\#2} & \textbf{AWG\#3} \\
        \hline
        \textbf{Material platform} & \(\text{Si}_3\text{N}_4\) & \(\text{Si}_3\text{N}_4\) & Doped-\(\text{SiO}_2\) \\
        \hline
        \textbf{Waveguide Geometry} & 1000×200 nm & 1000×200 nm & 3400×3400 nm \\
        \hline
        \textbf{Effective index (TE)} & 1.58 & 1.58 & 1.49 \\
        \textbf{Effective index (TM)} & 1.50 & 1.50 & 1.49 \\
        \hline
        \textbf{Min. \(R_{\text{curve}}\)} & \textit{500 \(\mu m\)} & \textit{500 \(\mu m\)} & \textit{1500 \(\mu m\)} \\
        \hline
        \textbf{Channel Spacing (\(\Delta \lambda\))} & 7.5 nm & 8 nm & 8.75 nm \\
        \hline
        \textbf{Central wavelength} & 1550 nm & 1550 nm & 1550 nm \\
        \hline
        \textbf{Resolving power (\(\lambda / \Delta \lambda\))} & 200 & 190 & 175 \\
        \hline
        \textbf{FSR} & 180 nm & 350 nm & 200 nm \\
        \hline
        \textbf{Footprint} & 11.75×5.2 mm\(^2\) & 11.75×9.2 mm\(^2\) & 11×3.5 mm\(^2\) \\
        \hline
    \end{tabular}
    \caption{Comparison of different AWG designs. Table based on \textit{P. Gatkine et al.} \cite{1}.}
    \label{table1}
\end{table}

\begin{figure}[htbp]
\centering % Centers the image
\includegraphics[width=0.45\textwidth]{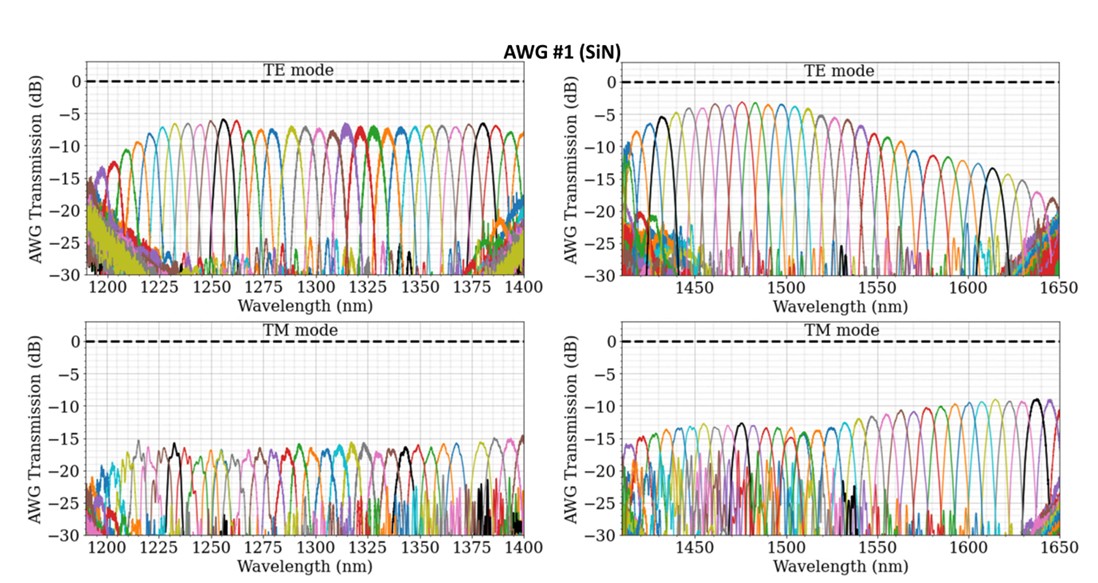} % Make sure the file path is correct
\caption{Transmission response of AWG\#1 in the TE and TM modes in the 1200 to 1600 nm range. Adapted from \textit{P. Gatkine et al.}\cite{1}.}
\label{fig7}
\end{figure}

\begin{figure}[htbp]
\centering % Centers the image
\includegraphics[width=0.45\textwidth]{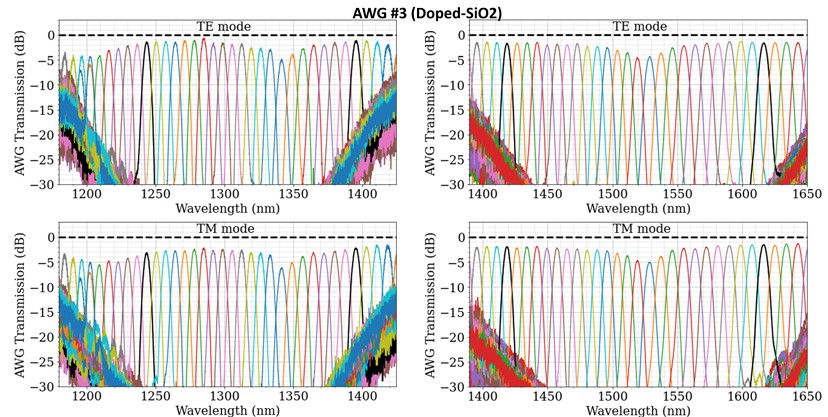}
\caption{Transmission response of AWG\#3 in the TE and TM modes in the 1200 to 1600 nm range. Adapted from \textit{P. Gatkine et al.}\cite{1}.
}
\label{fig8}
\end{figure}

Based on their result, they achieved the highest resolving power for their AWG on a \(\text{Si}_3\text{N}_4\) platform with a 1000x200 nm waveguide dimension at a central wavelength of 1550 nm. However, from Fig. \ref{fig7} and Fig. \ref{fig8}. We can see a significant loss of transmission for AWG\#1, which is also wavelength and polarization dependent.

From Fig. \ref{fig7} we can see that peak transmission was obtained at around 1450 nm for Transverse Electric (TE) mode, while in Transverse Magnetic (TM) mode a gradual increase can be observed with increasing wavelength. There is also a rapid decline at higher wavelengths for the TE mode.

\begin{figure}[htbp]
\centering % Centers the image
\includegraphics[width=0.5\textwidth]{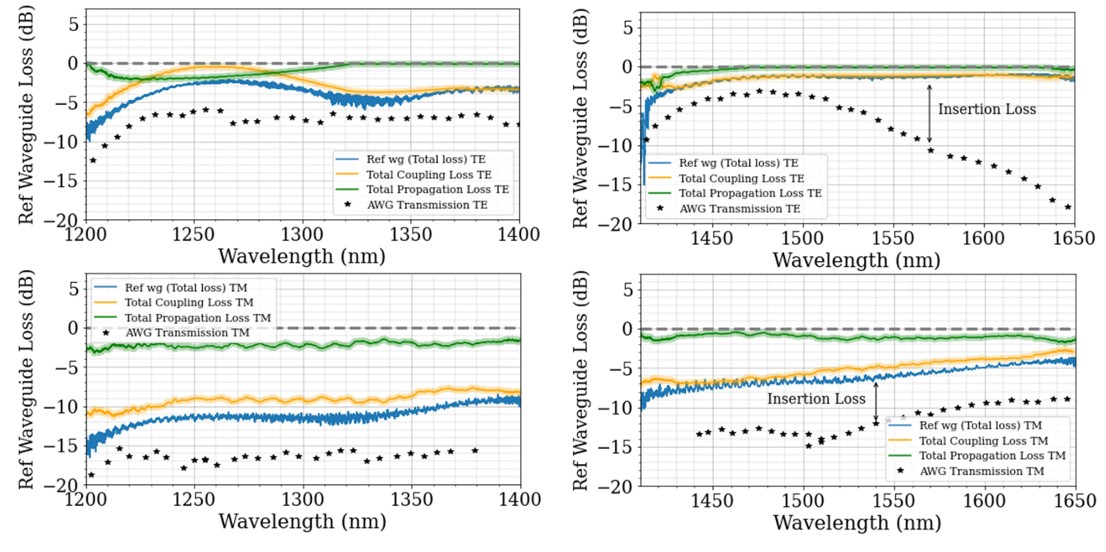}
\caption{Transmission response in TE and TM modes for AWG\#1 along with loss components. Adapted from \textit{P. Gatkine et al.}\cite{1}.
}
\label{fig9}
\end{figure}

The loss components are plotted against wavelength in Fig. \ref{fig9} separately for TE and TM modes. The plot also shows the transmission in these modes. Two reference waveguides were also used in this analysis. The transmission loss is reported to be mainly caused by the coupling loss between the fiber and waveguide. On the other hand, the AWG\#3 on a doped \(\text{SiO}_2\) platform operates over the same wavelength range without a significant drop in transmission for either polarization mode. Although the resolving power is slightly lower than AWG\#1, the study suggested that AWG\#3 was an ideal option for low-resolution, broadband on-chip spectroscopy. This conclusion was based on its fiber-to-chip coupling loss, broadband low-loss performance, low crosstalk, and insensitivity to polarization.

\subsection{Spectral Calibration}

\begin{figure}[htbp]
\centering % Centers the image
\includegraphics[width=0.45\textwidth]{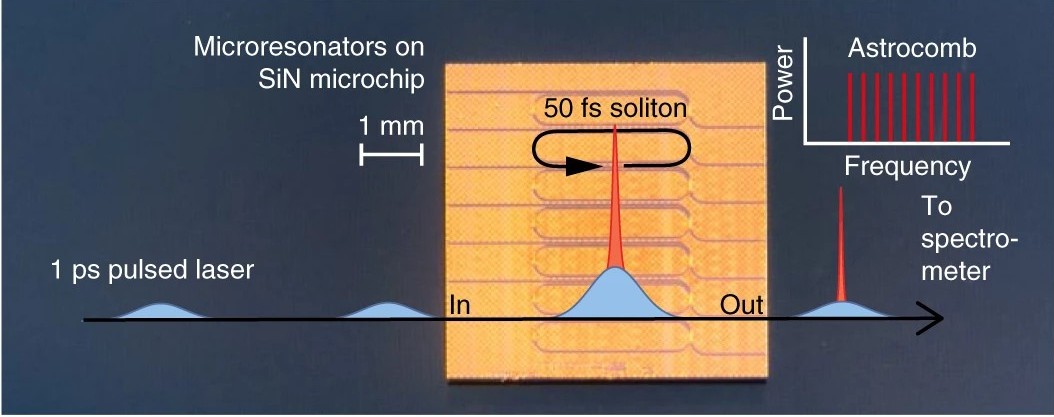}
\caption{Working concept of the MRR astrocomb \cite{10}.
}
\label{fig10}
\end{figure}

Detecting a slight wavelength shift in the optical spectrum of astronomical objects requires extremely accurate spectrometer calibration. This can be achieved through the use of laser frequency combs. The study carried out by Obrzud et al. demonstrates a microphotonic astrocomb used for this purpose \cite{10}. The design mainly consists of a silicon nitride-based microring resonator (MRR) structure on a chip \cite{11}. Their device exhibits a free spectral range (FSR) of 23.7~GHz, a Q-factor of \(6.4 \times 10^5\), and a finesse of 79. Fig. \ref{fig10} shows the working process of the astrocomb. Another key feature of this approach is that no spectral filtering is required. Moreover, the comb also generates a spurious-free broadband spectrum through the use of dissipative Kerr solitons in the MRR. This helps achieve the large FSR even without Fabry-Perot Bragg (FPB) filtering. Lastly, the on-chip comb allows for compact integration in astronomical applications. Stabilization to atomic frequency standards for precise wavelength calibration is crucial for the detection of exoplanets.

\subsection{High-Contrast Imaging}

High-contrast observation is a technique used to detect faint objects near bright stars. The survey conducted by Oppenheimer et al. on advancements in high-contrast optical and infrared astronomy explores methods such as adaptive optics (AO), coronagraphy, interferometry, and speckle suppression \cite{12}. As discussed in the previous section, AO is used to generate diffraction-limited images from telescopes. In this case, AO corrects distortions, such as stabilizing the image of a star using a tip/tilt system. Furthermore, wavefront correction is achieved with a deformable mirror. These mirrors have hundreds or thousands of actuators beneath a thin reflective layer and can correct up to a few microns of phase error.

\begin{figure}[htbp]
\centering % Centers the image
\includegraphics[width=0.48\textwidth]{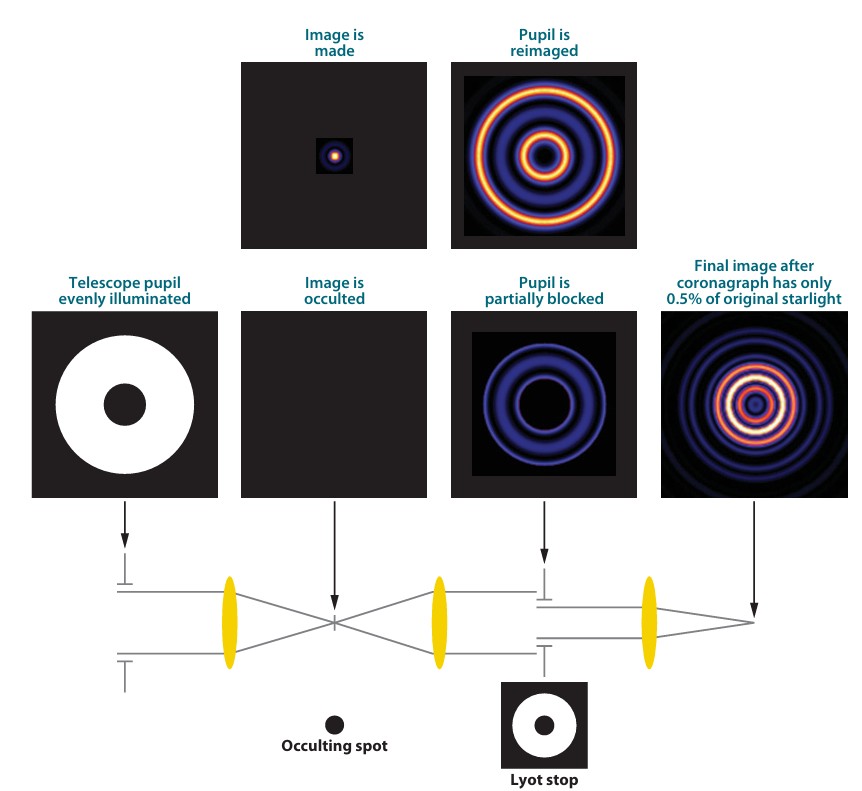}
\caption{Workflow of the Lyot Coronography \cite{13}.
}
\label{fig11}
\end{figure}

The process of coronagraphy is illustrated in Fig. \ref{fig11}. By strategically using masks and phase manipulation, the coronagraph can block most of the starlight.

Interferometry has seen significant use in astronomical applications, but there is still much to explore regarding interferometers in the context of high-contrast imaging. The potential applications are tremendous in this regard. N. Jovanovic et al. conducted a study on an integrated photonic pupil-remapping interferometer for high-contrast imaging \cite{14}. Its potential for exoplanet detection, when paired with adaptive optics systems, was validated by laboratory and on-sky tests conducted at the Anglo-Australian Telescope, which demonstrated stable performance and high-visibility fringes.

One of the main problems with high-contrast imaging is the uncontrolled amount of wavefront error and the need to suppress speckles. Studies like that carried out by G. Deng et al. shows how mordern machine learning models can be introduced into the denoising algorithm that can greatly increase spectral reconstruction accuracy compared to traditional methods\cite{19}. Their study employs long short-term memory (LSTM) for this case. LSTM has been used in various applications related to spectrometry and also for other purposes\cite{20}. Table \ref{table2} compares the different wavefront qualities required for various levels of contrast. To suppress speckles, data sensitive to speckles must be processed properly so that speckles can be distinguished from real sources and removed. 

\begin{table}[h!]
\centering
\renewcommand{\arraystretch}{1} % Adjust row height
\setlength{\tabcolsep}{2pt} % Adjust column spacing
\begin{tabular}{|l|c|c|c|c|c|}
\hline
\textbf{Contrast} & \textbf{\begin{tabular}[c]{@{}c@{}}Coherent \\ Wave-Front \\ Error ($\lambda$)\end{tabular}} & \textbf{\begin{tabular}[c]{@{}c@{}}RMS\\ Wave-Front \\ Error ($\lambda$)\end{tabular}} & \textbf{\begin{tabular}[c]{@{}c@{}}RMS\\ Path-Length \\ Error (nm)\end{tabular}} & \textbf{\begin{tabular}[c]{@{}c@{}}Reduced \\ Coherence \\Time (ms)\end{tabular}} & \textbf{\begin{tabular}[c]{@{}c@{}}Guidestar \\ H Magnitude\end{tabular}} \\ \hline
$10^6$  & $\lambda / 4400$  & $\lambda / 88$   & 18.7  & 1.07  & 6.9   \\ \hline
$10^7$  & $\lambda / 14000$ & $\lambda / 280$  & 5.7   & 0.34  & 3.2   \\ \hline
$10^8$  & $\lambda / 44000$ & $\lambda / 880$  & 1.9   & 0.11  & -0.6  \\ \hline
$10^9$  & $\lambda / 140000$& $\lambda / 2800$ & 0.6   & 0.03  & -4.3  \\ \hline
$10^{10}$& $\lambda / 440000$& $\lambda / 8800$& 0.2   & 0.01  & -8.1  \\ \hline
\end{tabular}
\caption{Maximum wave-front error for a given contrast \cite{15}.}
\label{table2}
\end{table}

\section{Future Work}

 Ongoing research across various areas of astrophotonics is leading to the discovery of low-loss and compact devices that have the potential to be used in astronomical studies. With significant strides made in fields such as interferometry, astrophotonic spectrography, and spectral calibration, this area of research is undoubtedly of great interest and holds immense potential. In this paper, we have reviewed four broad areas of photonic integration in astronomical instruments. With improvements in wavefront and distortion correction methods, along with the technological ability to process data to generate high-contrast images, astrophotonics will, very soon, significantly aid astronomers in the direct imaging of exoplanets as well as in the study of planetary and galaxy formations.
%Citations%

\bibliographystyle{IEEEtran}
\bibliography{References.bib}

\end{document}